\documentclass[rmp,aps,showpacs,floats,superscriptaddress,floatfix]{revtex4-2}
\usepackage{amssymb}
\usepackage{graphicx}
\usepackage{bm}
\usepackage{amsfonts}
\usepackage{color}
\usepackage{amsmath}    
\usepackage{epsfig}
\usepackage{hyperref}   
\usepackage{mathrsfs}
\usepackage{enumitem}

\usepackage{xcolor}
\hypersetup{
    colorlinks,
    linkcolor={red!50!black},
    citecolor={blue!50!black},
    urlcolor={blue!80!black}
}

\renewcommand{\index}[1]{#1}  

\newcommand{\matr}[1]{\mathbf{#1}}     

\begin{document}
\title{Centralities in complex networks}
\author{Alexandre Bovet}
\email{alexandre.bovet@maths.ox.ac.uk}
\affiliation{Mathematical Institute, University of Oxford, United Kingdom}
\author{Hern\'an A. Makse}
\email{hmakse@ccny.cuny.edu}
\affiliation{Levich Institute and Physics Department, City College of New York, New York, NY 10031, USA}

\maketitle

\section{Glossary}

\begin{itemize}[leftmargin=3cm]
    \item[Network]  A \emph{network} is a collection of nodes (also called vertices) and edges (also called links) linking pair of nodes. Mathematically, it is represented by a graph ${G}=({V},{E})$ where ${V}$ is the set of nodes and ${E}\subseteq V\times V$ is the set of edges. Additional information can be attached to each node or edge, for example edges can have different weights. Edges can be undirected or directed.
    
    \item[Adjacency matrix] The \emph{adjacency matrix},  $\matr{A}$,  of a network is a $N\times N$ matrix ($N=\left|{V}\right|$) with element $A_{ij}=1$ if there is an edge from node $i$ and to node $j$ and $A_{ij}=0$ otherwise. If the network is weighted, $A_{ij}=w_{ij}$ where $w_{ij}\in\mathbb{R}$ is the weight associated with the edge between nodes $i$ and $j$ if it exists and $A_{ij}=0$ otherwise.
    For undirected network, $A_{ij}=A_{ji}$, i.e. $\matr{A}$ is symmetric.
    
    \item[Degree of a node]  The \emph{degree}, $k_i$, of node $i$ in an undirected network is equal to its number of connections, i.e. $k_i=\sum_j A_{ij}$. For directed network, we differentiate the \emph{in-degree}, $k^{in}_i=\sum_j A_{ji}$ and the \emph{out-degree}, $k^{out}_i=\sum_j A_{ij}$, i.e. the number of edges in-coming to node $i$ and out-coming from node $i$, respectively. In weighted undirected networks, the degree of a node is replaced by its \emph{strength}, $s_i=\sum_j w_{ij}$ or \emph{in-strength} and \emph{out-strength} for weighted directed networks.
    
    \item[Walk] A \emph{walk} is an alternating sequence of vertices and edges in which every vertex is incident to both the edges that come before and after it in the sequence.
    \item[Path] A \emph{path} on a graph is a walk in which all vertices and edges are distinct.
    \item[Connected components] A \emph{connected component} of an undirected graph $G(V,E)$ is a subgraph of $G$, made of a subset of $V$ and all the edges connecting nodes of the subset toghether,  where there exist a path between each pair of nodes. In directed graphs, we differentiate \emph{strongly-connected components}, where there exist a path in both directions between all pairs of nodes, and \emph{weakly-connected components}, where there exist a path in at least one direction between all pairs of nodes.
\end{itemize}

\section{Why study networks?}

In \index{network science} complex systems are represented as a mathematical \index{graphs} consisting of a set of nodes representing the components and a set of edges representing their interactions.
The framework of networks has led to significant advances in the understanding of the 
structure, formation and function of complex systems\,\cite{newman2003structure,barrat2008dynamical,boccaletti2006complex,albert2002statistical}.
Social and biological processes 
such as the dynamics of epidemics\,\cite{pastor2015epidemic}, 
the diffusion of information in social media\,\cite{zhang2016dynamics},
the interactions between species in ecosystems\,\cite{montoya2006ecological} or the  communication between neurons in our brains\,\cite{bullmore2009complex} are all actively studied using dynamical models on complex networks.
In all of these systems, the patterns of connections at the individual level play a fundamental role on the global dynamics and finding the most important nodes allows one to better understand and predict their behaviors.

\section{Definition of the Subject}
Real-world \index{complex networks} are characterized by a number of structural features differentiating them from regular networks, such as lattices, but also making them different than completely random graphs, such as Erd\H{o}s-R\'enyi graphs\,\cite{newman2003structure}.
The properties that can be found in complex networks include, for example, a modular organization, also called community structure\,\cite{Fortunato2010}, or the so-called small-world effect, i.e. the fact that most pairs of nodes are connected by a very short path compared to the sizes of the networks\,\cite{Watts1998}.
A characteristic of real-world complex networks that interests us here and that has been intensively studied is the fact that their degree distributions are usually very heterogeneous\,\cite{albert2002statistical,newman2003structure}, indicating that there is usually large differences in the number of connections that nodes have.
Many real world networks have been found to have degree distributions resembling power laws and are sometimes referred to as scale-free\,\cite{barabasi2009scale}.
Whether these real-world systems are really scale-free and whether their degree distributions are really power laws is still disputed\,\cite{Holme2019}.
However, what is undeniable is the fact that many real world networks have degree distributions that are heavy-tailed with a minority of the nodes concentrating the majority of the connections.

In these systems, a small set of essential nodes can shape the collective dynamics of the entire systems.
For example, during epidemic outbreaks of infectious diseases, some individuals, known as \emph{super-spreaders}, infect disproportionately more secondary contacts, as compared to most others\,\cite{kitsak2010identification,stein2011super},
keystone species in ecology are responsible for the integrity and stability of ecosystems\,\cite{may1972will,scheffer2012anticipating,mills1993keystone,morone2018kcore}
and specific regions in brain networks are more important than others in the formation of memory\,\cite{bullmore2009complex,del2018finding,reis2014avoiding,zamora2010cortical}.
In social networks, a small set of \index{influencers} can drive the global dynamics of the system\,\cite{bovet2019influence} and opinion leaders are capable of influencing the public viewpoint on certain trending topics \,\cite{watts2007influentials}.
An important research effort in network science has therefore been dedicated to the development of methods allowing to find the most important nodes in networks. 
Intuitively, nodes with a large degree are likely to be more successful to trigger large-scale propagations or to control a large number of nodes. The degree centrality ranks nodes in terms of their degree and allows to identify highly connected \emph{\index{hubs}} present in most real-world complex networks that play an essential role in controlling their dynamics and maintaining their integrity\,\cite{barabasi1999emergence,albert2000error,cohen2001breakdown}.
While the \index{degree centrality} is arguably the simplest \index{centrality measure}, it only uses local information about each node to rank its centrality and more complex centrality measures have been developed in order to capture the collective network effects impacting the influence of a node.
Most centrality measures are based on a notion of distance between nodes or on a way to traverse the network that allows to compute how "central" a node is compared to other nodes while capturing the heterogeneous structural patterns of complex networks.
In the following, we describe centrality measures based on the notions of \index{network traversal} they rely on.
This short entry aims at being an introduction to this extremely vast topic, with many contributions from several fields, and is by no means an exhaustive review of all the literature about network centralities (see, for example, \,\cite{freeman1978centrality,Borgatti2005,Perra2008,landherr2010critical,rodrigues2019network} for more exhaustive reviews).

\subsection{Centrality measures based on shortest paths}

\index{Closeness centrality}\,\cite{bavelas1950communication, beauchamp1965improved,sabidussi1966centrality,dangalchev2006residual} and \index{betweenness centrality}\,\cite{freeman1978centrality,friedkin1991theoretical} are two dual measures of centrality initially developed in social sciences to assess the importance in terms of ease of access to others nodes (closeness) and brokering power (betweenness) of individuals in social networks.
These centralities have since then been used in many other contexts.
They are based on the assumption that information, or influence, propagates between two nodes in the most efficient way, i.e. by following the shortest path between them.

Considering an unweighted and undirected graph $G=(V,E)$, a walk on the graph $G$ is an alternating sequence of vertices and edges in which every vertex is incident to both the edges that come before and after it in the sequence. A path on a graph is a walk in which all vertices and edges are distinct.
A graph is said to be connected if there exists a path from any node node to any other node in the graph.
Considering the distance $\text{dist}(s,r)$ between two nodes $s$ and $r$ in a connected undirected and unweighted graph $G=(V,E)$ as the number of edges in the \index{shortest path} between them, the total distance of vertex $v$ is defined as the sum of its distance to all other vertices

$$\text{dist}(s) = \sum_{r\in V} \text{dist}(s,r),$$

which is larger for vertices that are the farther away from other vertices.
Therefore, the \emph{closeness centrality} of a node $s$ on a connected and undirected network is usually defined as\,\cite{bavelas1950communication, beauchamp1965improved}

\begin{equation}
    c_C(s) = \frac{1}{\sum_{r\in V} \text{dist}(s,r)} = \frac{1}{\text{dist}(s)}.
\end{equation}

The betweenness centrality captures the brokering power of node as the opportunity it has to intercept or influence the communications happening between pairs of other nodes. 
Let $\sigma(s,r)$ be the number of shortest paths between $s$ and $r$ and let $\sigma(s,r|b)$ be the number of shortest paths between $s$ and $r$ passing by a brokering node $b\in V\setminus \{s,r\}$. We consider $\sigma(s,s)=1$ and $\sigma(s,r|b)=0$ if $b\in\{s,r\}$.
We call the quantity $\delta(s,b,r)=\frac{\sigma(s,r|b)}{\sigma(s,r)}$ the \emph{dependency} of a sender $s$ and a receiver $r$ on a broker $b$\,\cite{Brandes2016}.
The \emph{betweenness centrality} in a connected and undirected graph $G=(V,E)$ is defined as the sum of dependencies of all communicating pairs on a broker $b$\,\cite{freeman1980gatekeeper}

\begin{equation}
    c_B(b) = \sum_{s,r\in V} \frac{\sigma(s,r|b)}{\sigma(s,r)}= \sum_{s,r\in V} \delta(s,b,r),
    \label{eq:betweenness}
\end{equation}

and can be seen as the overall potential control of $b$ on the communications in G.

\begin{figure}[ht]
\centering
 \includegraphics[width=0.5\linewidth]{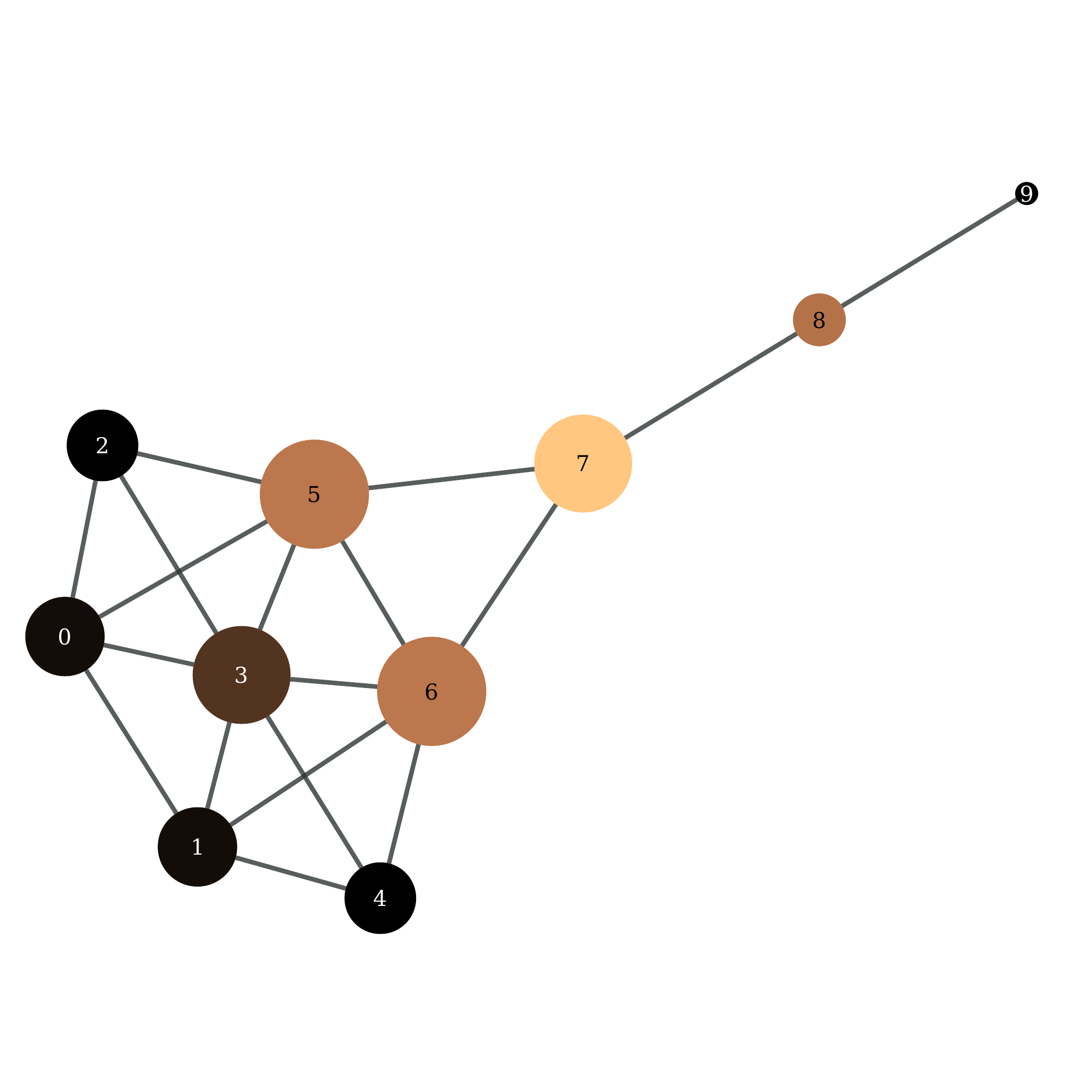}
 \caption{Krackhardt kite graph showing the different ranking obtained using degree, closeness and betweenness centralities\,\cite{krackhardt1990assessing}. 
 The closeness of a node is represented by its size and the betweenness of a node is indicated by its color with nodes with larger betweenness being of a lighter shade.
 The node with largest degree centrality is 3.
 Nodes 5 and 6 have the largest closeness centrality and node 7 has the largest betweenness centrality.
 }
 \label{fig:closness_vs_betweenness}
\end{figure}

Betweenness and closeness centrality can be seen as being dual to each other conceptually expressing either the independence from the control of others (closeness) or the potential control over others (betweenness)\,\cite{freeman1980gatekeeper,Brandes2016}.
Indeed, by noting that when considering shortest paths between $s$ and $r$ with different brokers $b$ at a fixed distance $d$ from $s$ ($1 \leq d < \text{dist}(s,r)$) each shortest path pass through exactly one broker and therefore 
$\sum_{b\in V : \text{dist}(s,b)=d} \frac{\sigma(s,r|b)}{\sigma(s,r)} = 1$.
This implies that the sum of dependencies between a sender $s$ and a receiver $r$ taken over of possible brokers $b$ is proportional to the distance between $s$ and $r$: $\sum_b \delta(s,b,r)= \sum_{d=1}^{\text{dist}(s,r)-1} \sum_{b\in V : \text{dist}(s,b)=d} \frac{\sigma(s,r|b)}{\sigma(s,r)} =\text{dist}(s,r)-1$.
This observation allows one to define the closeness centrality as $c_C(s)^{-1} = (N-1) + \sum_{b,r\in{V}}\delta(s,b,r)$ revealing its mathematical duality with betweenness centrality as a different partial sum, over $b$ and $r$ instead of $s$ and $r$, of the dependencies $\delta(s,b,r)$ showing its interpretation as a measure of lack of independence on others\,\cite{Brandes2016}.
Figure \ref{fig:closness_vs_betweenness} shows the kite graph introduced by  Krackhardt in 1990 as an illustration of the different ranking obtained by these centrality measures\,\cite{krackhardt1990assessing}.

In directed networks, closeness and betweenness centrality can be generalized by considering the notion of reachability instead of distance. In weighted networks where edge weights typically represent a distance or a lag in the connection between adjacent nodes, the length of a path is usually taken as being the sum of the weights of its edges.
We refer the reader to \,\cite{Brandes2016} for a discussion about the generalizations of the closeness and betweenness centrality to directed and weighted networks, and to their interpretations in these cases.

Many variants of closeness and betweenness have been developed\,\cite{brandes2008variants}. The concept of betweenness centrality has also been extended to edges by considering the number of shortest paths containing an edge instead of a node in eq. (\ref{eq:betweenness})\,\cite{brandes2008variants} or, for example, by considering the fraction of minimum spanning trees of a graph that contain a given edge\,\cite{teixeira2015not}.
Versions of the closeness and betweenness centralities based on random walks instead of shortest paths have also been developed\,\cite{white2003algorithms,Newman2005}.

\subsection{Centrality measures based on walks}

Several widely used centrality measures can be seen as being based on the concept of \index{graph walks}. 
Walks, alternating sequences of vertices and edges in which every vertex is incident to both the edges that come before and after it in the sequence, are useful to count the number of possible ways there is to reach a given node starting from another node.
Centrality measures based on walks usually try to find the nodes from which there are the largest number of walks reaching other nodes.
For unweighted networks, the number of walks of length $\ell$ existing between two nodes $i$ and $j$ is given by the element $(i,j)$ of the $l^{\text{th}}$ power of the adjacency matrix: $\left(\matr{A}^\ell\right)_{ij}$.
The number of walks of length $\ell$ starting from node $i$ is then given by $w(\ell)_i=\sum_j\left(\matr{A}^\ell\right)_{ij}$ or in matrix notation $\matr{w}(\ell)=\matr{A}^\ell \matr{1}$, where $\matr{1}$ is the unit vector.
The \emph{degree centrality} of a node $i$ can be expressed as the number of walks of length 1 starting from it:
\begin{equation}
    c_{\text{deg}}(i) = w(1)_i = \sum_j A_{ij}.
\end{equation}

In order to take into account information about the surrounding of a node in its centrality score, several researchers have developed centrality measures that consider longer walks. The idea being that if a node is close to a node with a high centrality, its centrality should also be high\,\cite{katz1953new,bonacich1972factoring}.
Starting from an initial guess for the centrality of each node given by the vector $\matr{x}(0)$ (e.g. $\matr{x}(0)=\matr{1}$), we can propagate this initial centrality through walks of a given length. 
The centrality vector for walks of length $\ell$ is given by 
\begin{equation}
    \matr{x}(\ell) = \matr{A}^\ell \matr{x}(0).
    \label{eq:eig_cen_it}
\end{equation}

Writing the initial centrality as a linear combination of the eigenvector of the adjacency matrix, one can see that as $\ell\xrightarrow{}\infty$ the centrality will converge to a vector proportional to the leading eigenvector of $\matr{A}$, the one corresponding to its largest eigenvalue, and that takes into account global information about the network structure\,\cite{Newman2010networks}.
As $\matr{A}$ is non-negative and if $G$ is connected, the \index{Perron-Frobenius theorem} ensures that the leading eigenvector is positive and that the associated eigenvalue is positive and simple.
The resulting vector is called the \emph{\index{eigenvector centrality}} whose element $i$ is defined as 

\begin{equation}
    c_\text{eig}(i) \propto \sum_j A_{ij} c_\text{eig}(j).
\end{equation}

\begin{figure}[ht]
\centering
 \includegraphics[width=0.8\linewidth]{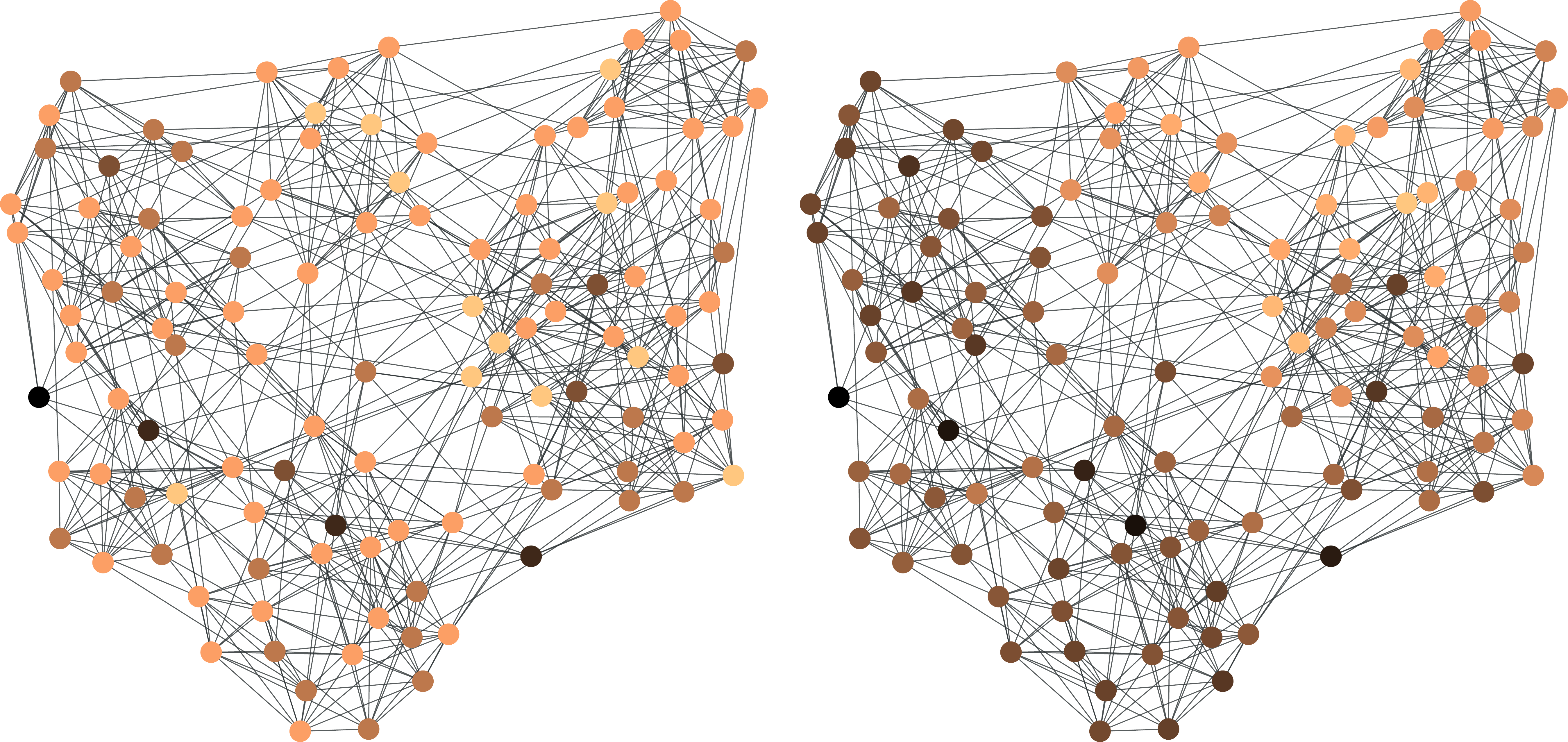}
 \caption{Network of American football games between Division IA colleges during regular season Fall 2000\,\cite{girvan2002community,evans2010clique}.
 Degree centrality for this undirected graph is shown on the left and eigenvector centrality is shown on the right. Lighter shades indicate a higher centrality.
 Several nodes have the same degree in this network making the degree centrality unable to disantangle the importance of nodes with the same degree.
 Eigenvector centrality has a tendency of being localized in regions where many high degree nodes are close to each other.
 }
 \label{fig:deg_vs_eigen}
\end{figure}

In weighted networks, the eigenvector centrality is propagated along walks on the graph that are weighted by the edge weights.
The eigenvector centrality can be computed similarly on directed and undirected networks with the difference that for directed networks, the adjacency matrix is in general asymmetric and has therefore two different sets of left- and right-eigenvectors. Depending on the application, one may prefer to use the largest left-eigenvector or the largest right-eigenvector if one is more interested in nodes having a large number of incoming walks or outgoing walks, respectively.
However, other types of centralities are usually preferred for directed networks as the eigenvector centrality of nodes outside of a strongly connected component (a subgraph where there exist a path in both directions between all pairs of nodes) may be equal to zero. In particular, in directed graphs without cycles (closed directed paths), the eigenvector centrality is zero for all nodes.
A comparison of the degree centrality with the eigenvector centrality is shown in Fig. \ref{fig:deg_vs_eigen} for an undirected network of American football teams\,\cite{girvan2002community,evans2010clique}.

The \emph{\index{subgraph centrality}} is based on the idea of counting the number of times that a node takes part in the different connected subgraphs of a network, with smaller subgraphs having higher importance\,\cite{Estrada2005}.
The subgraphs are identified by counting closed walks of different lengths that start and end on a given node. The subgraph centrality of node $i$ is defined as

\begin{equation}
    c_{\text{Sub}}(i) = \sum_{\ell=0}^{\infty}\frac{\left(\matr{A}^\ell\right)_{ii}}{\ell!}.
\end{equation}

By noticing that $c_{\text{Sub}}(i)=\left(e^\matr{A}\right)_{ii}$, where $e^\matr{A}$ denotes the matrix exponential of $\matr{A}$,
the subgraph centrality can be obtained from the spectrum of the adjacency matrix as\,\cite{Estrada2005}

\begin{equation}
    c_{\text{Sub}}(i) = \sum_{i=1}^{N}(v_j^i)^2e^{\lambda_j}, 
\end{equation}

where $v_1,v_2, \ldots, v_N$ form an orthonormal basis of $\mathbb{R}^N$ and are eigenvectors of $\matr{A}$ associated with the eigenvalues $\lambda_1, \lambda_2, \ldots, \lambda_N$.
This centrality is more discriminative than the degree, betweenness, closeness or eigenvector centralities and gives a distinctly different ranking of nodes in real-world complex networks\,\cite{Estrada2005}.

Another centrality measures based on walks that tries to solve the issue of the eigenvector centrality in directed networks is the \index{Katz centrality}\,\cite{katz1953new}.
The Katz centrality of node $i$ is the weighted sum of all walks emanating from $i$, with the count for walks of length $\ell$ weighted by a factor $\alpha^\ell$ where $0<\alpha<1$. In this way the importance of longer walks is diminished compared to shorter walks.
The \emph{Katz centrality} of node $i$ can be written as\,\cite{katz1953new}

\begin{equation}
    c_{\text{Katz}}(i) = 1 + \sum_{\ell=1}^{\infty}\sum_{j} \alpha^\ell \left(\matr{A}^\ell\right)_{ij},
\end{equation}

where the unit shift does not alter the ranking of the nodes and may be seen as arising from a single walk of length zero.
In order for the centrality to converge, the factor $\alpha$ must be smaller than the reciprocal of the absolute value of the largest eigenvalue of $\matr{A}$\,\cite{katz1953new,bonacich1972factoring,bonacich1987power}.
In this case, we can also note that $\matr{I}+\alpha\matr{A}+\alpha^2\matr{A}^2+\alpha^3\matr{A}^3 + \cdots = \left(\matr{I}-\alpha\matr{A}\right)^{-1}$.
Thus, the vector $\matr{x}$ giving the Katz centrality of each node is also solution of the matrix equation $\matr{x}=\left(\matr{I}-\alpha\matr{A}\right)^{-1}\matr{1}$ which can be rewritten as $\matr{x}=\alpha\matr{A}\matr{x}+\matr{1}$.
The Katz centrality can therefore be computed iteratively in a similar manner than the eigenvector centrality (with eq. (\ref{eq:eig_cen_it})) using 
\begin{equation}
    \matr{x}(t+1) = \alpha\matr{A}\matr{x}(t)+1.
    \label{eq:katz_it}
\end{equation}

A unit value is added at each iteration guaranteeing that the Katz centrality will never be equal to zero, even in directed networks.
The factor $\alpha$ can also be seen as balancing the importance of the eigenvector term compared to the constant term added at each iteration.
In the limit as $\alpha$ approaches the reciprocal of the absolute value of the largest eigenvalue of $\matr{A}$, the Katz centrality approaches the eigenvector centrality on strongly connected graphs\,\cite{bonacich1972factoring,bonacich1987power}.

Walk-based centralities have been generalized to hypergraphs (e.g. \,\cite{bonacich2004hyper,benson2019three}), multiplex networks (e.g. \,\cite{sola2013eigenvector,de2015ranking}) and temporal networks (e.g. \,\cite{nicosia2013graph,praprotnik2015spectral,taylor2017eigenvector}).

\subsection{Centrality measures based on random walks}

\begin{figure}[ht]
\centering
 \includegraphics[width=0.8\linewidth]{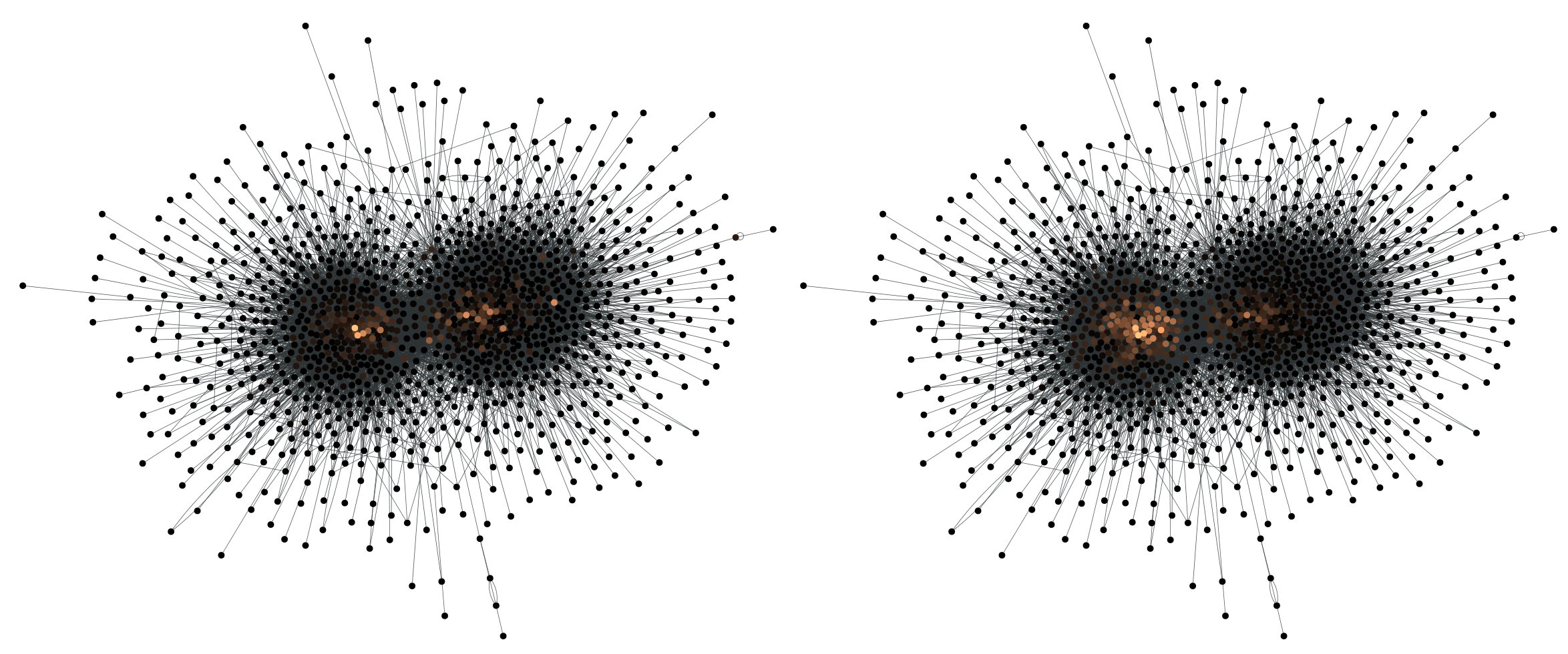}
 \caption{A directed network of hyperlinks between weblogs on US politics, recorded in 2005 by Adamic and Glance\,\cite{adamic2005political}.
 PageRank centrality is shown on the left and Katz centrality is shown on the right. Lighter shades indicate a higher centrality.
 The Katz centrality has a tendency of being "localized" in regions with many close-by hubs.
 }
 \label{fig:pr_vs_katz}
\end{figure}

\index{Random walks} on graphs have been extensively studied and have many applications in the study of diffusive processes on networks and for the characterization of the structure of complex networks \,\cite{Rosvall2008,Delvenne2010,durrett2010some,Masuda2016}.
A random walk on a graph is defined by the trajectory of a walker that, at each time step, jumps to a neighbors of $i$ with equal probability. On an unweighted and undirected network, the transition probability for a walker on node $i$ to jump to node $j$ is equal to 

\begin{equation}
    T_{ij} = \frac{A_{ij}}{k_i},
\end{equation}

or in matrix notation $\matr{T}=\matr{D}^{-1}\matr{A}$, where $\matr{D}=\text{diag}(\matr{k})$ is the diagonal degree matrix.
In directed networks, the degree is replaced by the out-degree and in weighted networks with positive weights, the probability transition is proportional to the weight of edge $(i,j)$.
Centrality measures based on random walks exploit the fact that,
if a centrality value propagates as a random walk, at each iteration, its value is divided across all out-going edges of a node.
This is different than for eigenvector and Katz centralities, based on walks, where the centrality tends to concentrate on a few hubs in the network, leaving other nodes almost indistinguishable with very low scores due to the repeated reflection of influence along the mutual connections between hubs during iterations of eqs. (\ref{eq:eig_cen_it}) \& (\ref{eq:katz_it}) (see Figs. \ref{fig:deg_vs_eigen} \& \ref{fig:pr_vs_katz}).

The distribution probability of finding a walker at time step $n$ on any node of the network can be written as a row-vector $\matr{p}(n)$ with $\sum_i p_i = 1$.
The evolution of the distribution probability is given by 

\begin{equation}
    \matr{p}(n+1) = \matr{p}(n)\matr{T}.
    \label{eq:rw_update}
\end{equation}

On connected undirected network, the random walk reaches a stationary distribution satisfying $\matr{p}^\star=\matr{p}^\star\matr{T}$ where the probability at each node is proportional to its degree:
\begin{equation}
    p^\star_i = \frac{k_i}{\sum_j k_j}.
\end{equation}

On undirected networks, the degree centrality can therefore also be seen as the stationary state of the random walk process.
On directed networks, random walks are not guaranteed to converge to a unique stationary state.
A stationary distribution only exists on strongly connected components of a directed network\,\cite{aldous2002reversible,Masuda2016}.

To overcome this issue the \emph{\index{PageRank}} centrality modifies the classical random walk by introducing a "teleportation" probability, i.e. at each step, the walkers have a given probability to teleport uniformly at random to any other nodes of the network. The update equation for the PageRank probability density is given by \,\cite{brin1998anatomy,page1999pagerank,Masuda2016}

\begin{equation}
    \matr{p}(n+1) = \alpha\matr{p}(n)\matr{S} + \frac{1-\alpha}{N}\matr{1}^T,
    \label{eq:pg_it}
\end{equation}

where $\matr{S}$ is a transition matrix constructed from $\matr{A}$ such that $S_{ij}=A_{ij}/k_\text{out}(i)$ when $k_\text{out}(j)>0$.
For "dangling nodes", that have no out-going edges (i.e. $k_\text{out}(i)=0$),
$S_{ij}=1/N$, meaning that a walker on such a node has a uniform probability to jump to any other nodes in the network.
On non-dangling nodes, at each step, walkers have a probability $\alpha<1$ to follow an out-going edge and a probability $1-\alpha$ to teleport to any node in the network.
This modified random walks is now ergodic also on networks that are not strongly connected and converges to the probability density vector giving the \emph{PageRank} centrality of node $i$ as

\begin{equation}
    c_\text{PR}(i) = \alpha \sum_j c_\text{PR}(j) S_{ji}  + \frac{1-\alpha}{N}
\end{equation}

which is equal to the normalized eigenvector corresponding to the largest positive eigenvalue of the so-called Google matrix $\matr{G}=\alpha \matr{S}+\frac{1-\alpha}{N}\matr{1}\matr{1}^T$\,\cite{ermann2015google}.
PageRank was famously developed for ranking websites for the search engine Google considering a "random surfer model" navigating through webpages by randomly clicking on hyperlinks\,\cite{page1999pagerank}.
PageRank also found many applications in other aspects, such as ranking scientists and academic papers \,\cite{ding2009pagerank}, images \,\cite{jing2008visualrank} and proteins \,\cite{ivan2010web}.
Figure \ref{fig:pr_vs_katz} shows a comparison of the PageRank and Katz centralities on a directed network of hyperlinks between weblogs\,\cite{adamic2005political}.

The similarities between equations (\ref{eq:eig_cen_it}) and (\ref{eq:rw_update}) as well as equations (\ref{eq:katz_it}) and (\ref{eq:pg_it}) reveal that the degree and the PageRank centralities can be seen as similar to the eigenvector and Katz centralities, but for random walks instead of regular graph walks. In the case of random walk based centralities, one looks for eigenvectors of transition matrices instead of adjacency matrices.

\subsection{Centrality measures based on non-backtracking walks}

In order to address the issue of "localization" of the eigenvector centrality mentioned above, i.e. when most of the weight of centrality vector concentrates around one or a few nodes in the network, several authors have focused on using \index{non-backtracking walks} instead of regular or random walks\,\cite{Martin2014,Arrigo2018,Arrigo2020}.
Non-backtracking walks are graph walks where backtracking steps, i.e. steps where the walk comes back to an immediately preceding node, are not permitted.
Using this type of walks can sometimes solve the problem of localization as they decrease reflections between hubs during iterations of the centrality\,\cite{Martin2014}.
Non-backtracking walks on unweighted networks are usually described using the $2M\times2M$ matrix $\matr{B}$ where rows and columns correspond to the directed edges of the network ($M=\left|{E}\right|$).
If the network is undirected, an equivalent directed network is considered by replacing each edge by a pair of directed edges in both directions.
Element ($i\rightarrow{}j,\ell\rightarrow{}h$) of $\matr{B}$ is equal to one only if $j=\ell$ and $i\ne h$, i.e. there is a non-backtracking path of length 2 $i\rightarrow{}j\rightarrow{}h$ in the network. 
More succinctly, 

\begin{equation}
    B_{i\rightarrow{}j,\ell\rightarrow{}h} = \delta_{j\ell}(1-\delta_{ih}),
\end{equation}

where $\delta_{ij}$ is the Kronecker delta.
The matrix $\matr{B}$ is referred to as the \emph{\index{non-backtracking matrix}} or the \emph{\index{Hashimoto matrix}}\,\cite{hashimoto1989zeta}.
Powers of $\matr{B}$ enumerate non-backtracking walks similarly to powers of the adjacency enumerate walks\,\cite{Arrigo2020}.
The non-backtracking matrix is in general asymmetric with eigenvalues that are in general complex, however, 
since its entries are non-negatives, by the \index{Perron-Frobenius theorem}, its leading eigenvalue, $\lambda$, is real and non-negative, and there exists a corresponding leading eigenvector, $\matr{v}$, whose elements are also non-negative real numbers.
If $G$ is connected and not a tree, i.e. it has at least one cycle, $\lambda$ is positive\,\cite{lin2019non}.
They satisfy the eigenvector equation

\begin{equation}
    \lambda\matr{v}=\matr{B}\matr{v}
\end{equation}

and the \emph{\index{non-backtracking eigenvector centrality}} of node $j$ is defined by \,\cite{Martin2014}

\begin{equation}
    c_\text{NBeig}(j) = \sum_i A_{ij}v_{i\rightarrow{}j}.
\end{equation}

Finding the leading eigenvector of $\matr{B}$ can however be computationally demanding for large graphs as the size of $\matr{B}$ is usually much larger than the size of $\matr{A}$.
However, the computation can be made much faster by computing directly $c_\text{NBeig}$ as the first $N$ elements of a $2N\times2N$ matrix called the Ihara-Bass matrix\,\cite{krzakala2013spectral}.

Non-backtracking walks have also been used to modify other centrality measures, such as the Katz\,\cite{Arrigo2018}, random walk\,\cite{lin2019non} or PageRank\,\cite{aleja2019non} centralities.
Research on centrality measures based on non-backtracking walks is very active with recent results showing that they can also suffer from localization issues on some networks\,\cite{barucca2016centrality,pastor2020localization}.\\

Non-backtracking walks are also used in the problem of \index{influence maximization}: finding the minimal set of nodes, the \emph{influencers},
which, if activated, would cause the spread of information to the whole network, or, if immunized, would prevent the diffusion
of a large scale epidemic\,\cite{kempe2003maximizing}.
Influence maximization can be mapped to the problem of \index{optimal percolation} of random networks\,\cite{morone2015influence} which consists in identifying the minimal set of nodes whose removal would dismember the network in many disconnected and non-extensive components. The fragmentation of the network is measured by the size of the largest connected component, called the giant component of the network.

The intuition behind the usage of non-backtracking walks in the optimal percolation problem comes from the fact that the giant component is held together by long paths and that powers of the non-backtracking matrix allows to quickly find them.
The removal of nodes is represented with the vector $\matr{n} =(n_1,\ldots,n_N)$ where $n_i =0$ if $i$ is removed (influencer) or $n_i =1$ otherwise.
Considering undirected locally tree-like random graphs,
a modified $2M\times 2M$ non-backtracking matrix $\matr{M}$ is then defined as 
$M_{i\rightarrow{}j,\ell\rightarrow{}h} = n_\ell B_{i\rightarrow{}j,\ell\rightarrow{}h}$.
Given an initial arbitrary positive vector $\matr{w}(0)$, repeated iterations with $\matr{M}$,

\begin{equation}
w(n)_{i\rightarrow{}j}=\sum_{kl}M_{i\rightarrow{}i,k\rightarrow{}l}w(n-1)_{k\rightarrow{}l},
\end{equation}

increase the norm of the vector as the influence of nodes on non-backtracking walks of larger and larger length 
are included.
The growth rate of the vector's norm is determined by the value of the largest eigenvalue, $\lambda(\matr{n})$ of the modified non-backtracking matrix.
As the influencer nodes are removed, the value of $\lambda(\matr{n})$ decreases until the giant component is reduced to a tree (a graph without cycles) plus only one cycle when $\lambda(\matr{n})$=1.
The removal of supplementary nodes quickly destroys the giant component and $\lambda(\matr{n})$ falls to zero\,\cite{morone2015influence}.
The optimal influence problem can therefore be rephrased as finding the optimal configuration $\matr{n}$ that minimizes the largest eigenvalue of $\matr{M}$\,\cite{morone2015influence}, which consists in removing the nodes that are on the most non-backtracking walks in the network and that are keeping the giant component connected.
The \emph{\index{collective influence}}\,\cite{morone2015influence} of a node, defined as 

\begin{equation}
    \text{CI}_\ell(i) = (k_i-1)\sum_{j\in \partial \text{Ball}(i,\ell)}(k_j - 1),
\end{equation}
where $\partial \text{Ball}(i,\ell)$ is the set of the nodes at a distance $\ell$ from $i$, is a measure of which node to remove in order to produce the largest diminution of the largest eigenvalue of non-backtracking matrix.
The ranking of the nodes in term of collective influence is produced by removing the node with the largest CI value, recomputing the CI values of its neighbors and repeating the operation until the giant component of the network disappears.
An advantage of CI is that it gives a high rank to seemingly "weak nodes" with a small number of connections that are in fact surrounded by highly connected nodes and therefore on the path of many non-backtracking walks.
The problem of influence maximization in networks is actively researched using many different approaches (see, for example, \,\cite{braunstein2016network,mugisha2016identifying,zdeborova2016fast,radicchi2016beyond,pan2016influence,pei2020influencer}).

\section*{Conclusion and Future Directions}

We have seen that there is not a single measure of centrality in networks and that one should carefully choose an appropriate measure depending on the subject of investigation. 
In practice, centrality measures should be chosen depending on whether the network is directed or not and whether, for example, one is looking for highly connected nodes, nodes with the most brokering power, nodes that share the most influence with each others or the minimal set of nodes that can spread information to the whole network. The differences in ranking obtained from different centrality measures should also be examined in order to better understand the signification of each ranking in a given context.

The research landscape on centrality measures for complex networks is much vaster than what has been presented in this short entry.
Research is very active with recent and future directions including, for example, the properties of centrality measures based on non-backtracking walks\,\cite{pastor2020localization},
centralities in signed networks (networks where weights can be positive or negative)\,\cite{liu2020simple} and 
multilayer networks\,\cite{sola2013eigenvector,de2015ranking,wang2018new,wu2019tensor}.
Research is also active on the development of centralities for specific applications based on network dynamics not necessarily captured by network traversals. Examples include centralities for social networks based on game theoretical concepts\,\cite{Shapley1953,Gomez2003,michalak2013efficient},  centralities for detecting vulnerabilities in power-grids based on networks of oscillators\,\cite{gutierrez2013vulnerability,tyloo2018robustness,Tyloo2019} or centralities for identifying key genes in gene regulatory networks based on the propagation of biological signals\,\cite{zotenko2008hubs,missiuro2009information,kim2011identifying,cowen2017network}.

Research is also ongoing about the development of centralities in network models that extend the concept of network as an ensemble of static pairwise relations, namely, centralities in temporal networks, i.e. time-evolving networks, \,\cite{nicosia2013graph,praprotnik2015spectral,taylor2017eigenvector,flores2018eigenvector,lv2019pagerank} and in hypergraphs or simplicial complexes, i.e. generalizations of graphs where an edge can join more than two nodes\,\cite{bonacich2004hyper,estrada2018centralities,benson2019three,aksoy2020hypernetwork,serrano2020centrality}.


%

\end{document}